\documentclass[a4paper,10pt, twocolumn]{article}
\usepackage{cite}
\usepackage{amsmath,amssymb,amsfonts}
\usepackage{algorithmic}
\usepackage{graphicx}
\usepackage{textcomp}
\usepackage{booktabs}
\usepackage{multirow}
\usepackage{xcolor}
\usepackage[version-1-compatibility]{siunitx}
\newcommand{\figH}{2.2in}

\date{}

\begin{document}

\title{Online Anomaly Detection in HPC Systems 
}

\author{\IEEEauthorblockN{Andrea Borghesi}
\IEEEauthorblockA{\textit{DISI}, \textit{University of Bologna}\\
Bologna, Italy\\
andrea.borghesi3@unibo.it}
\and
\IEEEauthorblockN{Antonio Libri}
\IEEEauthorblockA{\textit{IIS}, \textit{ETHZ}\\
Zurich, Switzerland \\
libria@iis.ee.ethz.ch}
\and
\IEEEauthorblockN{Luca Benini}
\IEEEauthorblockA{\textit{IIS}, \textit{ETHZ}\\
Zurich, Switzerland \\
lbenini@iis.ee.ethz.ch}
\and
\IEEEauthorblockN{Andrea Bartolini}
\IEEEauthorblockA{\textit{DEI}, \textit{University of Bologna}\\
Bologna, Italy \\
a.bartolini@unibo.it}
}

\author{Andrea Borghesi \\DISI, University of Bologna \and 
Antonio Libri \\Integrated Systems Laboratory, ETHZ 
\and Luca Benini \\Integrated Systems Laboratory, ETHZ
\and Andrea Bartolini \\DEI, University of Bologna}

\maketitle

\begin{abstract}
Reliability is a cumbersome problem in High Performance Computing Systems and
Data Centers evolution.  During operation, several types of fault conditions or
anomalies can arise, ranging from malfunctioning hardware to improper
configurations or imperfect software. Currently, system administrator and final
users have to discover it manually. Clearly this approach does not scale to
large scale supercomputers and facilities: automated methods to detect faults
and unhealthy conditions is needed.  Our method uses a type of neural network
called \emph{autoncoder} trained to learn the normal behavior of a real,
in-production HPC system and it is deployed on the edge of each computing node.
We obtain a very good accuracy (values ranging between 90\% and 95\%) and we
also demonstrate that the approach can be deployed on the supercomputer nodes
without negatively affecting the computing units performance.  
\end{abstract}

\section{Introduction}
\label{sec:intro}
Nowadays, supercomputers and large data centers are increasing in scale and
number of components, with systems composed of thousands\,/\,millions of
computing units\cite{Fu2016,exascaleProjections} and represent an increasingly
complex industrial plan.  Therefore, there is a huge number of sources of
possible faults, heterogeneous in their nature, ranging from hardware
malfunctions or misconfigurations, to software unwanted behaviours or bugs. For
system administrators who strive to guarantee systems operating in optimal
conditions, identifying faulty situations and anomalous behaviours is a daunting
task.

An automated online anomaly detection system capable to satisfy real time
requirements would be a boon for facility managers. Currently, most of the
state-of-the-art anomaly detection systems are based on the analysis of system
logs or log messages generated by dedicated software tools, often at OS
level\cite{oliner2007supercomputers,barth2008nagios,xu2010experience}. In this
way there is not a general and uniform detection system and, moreover, deploying
a set of different tools with different requirements still requires a lot of
effort by the system administrators. This fact curbs the number and types of
identifiable fault conditions with log-based tools. 

However, today's supercomputers and data centers have hardware components with
sensors to monitor physical and architectural parameters
\cite{beneventi2017continuous,DBLP:conf/cf/BartoliniBLBGTG18,LRZ,BartoliniUnveilingEurora}.
The integrated monitoring infrastructure periodically reads a set of metrics and
collects them into a single gathering point. The authors of
\cite{beneventi2017continuous} show that these sensors can easily reach 1.5KSa/s
per compute node, and propose Examon, a scalable infrastructure based on local
monitoring agents pushing data through the MQTT protocol. Clearly, local
software-based monitoring agents compete for the same computational resources of
users' applications. The authors of \cite{DiG,DBLP:conf/cf/BartoliniBLBGTG18}
propose out-of-band monitoring through edge computing devices, thus without
impacting users-dedicated computing resources. In this approach, an external
embedded device is inserted in the node, and it monitors the architectural and
physical sensors through a dedicated interface \cite{rosedahl2015chip}. 

It seems a sensible idea to use the collected data to look for possible
unhealthy situations. For example, a possible approach relies in
\emph{supervised} Machine Learning (ML) techniques, where a classifier is
trained to distinguish between healthy and abnormal behaviours. In recent years,
some approaches went in this direction, showing promising results but with a
somewhat limited scope. For instance, supervised methods need a carefully
prepared initial phase where the supercomputer is injected with all the kinds of
faults to be detected later. This is clearly a strong drawbacks because it does
not encompass the occurrence of new, unseen anomalies. 

In this paper we propose a novel automated method for anomaly detection in HPC
systems and data centers, based on a technique derived from the Machine Learning
(ML) area, namely a type of neural network called \emph{autoencoder}. The method
we propose has a very good accuracy (around 90\%-95\% of detection accuracy) and
can be directly executed in the edge, exploiting monitoring devices embedded
inside each computing node, guaranteeing real time performance and no overhead.
We demonstrate this approach on a real tier-1 supercomputer in production.

\section{Related Works}
\label{sec:related}
Tuncer et al. \cite{tuncer2017diagnosing} tackles the issue of diagnosing
performance variations in HPC systems. They collect several measurements through
a monitoring infrastructure; a group of statistical features modeling the state
of the supercomputer is obtained from these features. The authors then train
different ML algorithms to classify the behaviour of the supercomputer using the
statistical features. The results are promising, outperforming previous
state-of-the art\cite{bodik2010fingerprinting,lan2010toward}.  Baseman et al.
\cite{baseman2016interpretable} propose a similar technique for anomaly
detection in HPC systems. They apply a general statistical technique called
\emph{classifier-adjusted density estimation} (CADE) in order to help the
training of a supervised Random Forest classifier. The classifier decides the
class (normal, anomaly, etc) of each data point (set of physical measurements).

These methods both belong to the supervised area: the training set must contain
examples of all classes to be detected, e.g. examples of normal and abnormal
states, and must be unbiased (equal number of examples for each class). This
fact has a consequence: a first phase is required in order to create a labeled
data set and the supercomputer must be injected with faults. Furthermore,
supervised classifiers only learn to identify the classes already seen at
training time; unseen anomalies encountered at run time cannot be properly
detected by this methods. Our approach, thanks to the semi-supervised learning,
resolves both these issues.

Dani et al. \cite{dani2017k} describe instead an unsupervised technique for
anomaly detection in HPC. Their work is very different from our approach since
they consider only the console logs generated by computing nodes (no monitoring
infrastructure). Their purpose is to distinguish logs relative to faulty
conditions from logs created by healthy nodes; the proposed approach uses the
K-means clustering algorithm. Their work targets faults that can be recognized
by a node itself and stored in log messages; this bounds the number and the
types of detectable anomalies. Conversely, in our approach we detect anomalies
using the data gathered via a collection framework, without need for anomaly
detection systems already deployed on computing nodes.

\section{A Methodology for Automated Anomaly Detection}
\label{sec:approach}

\begin{figure}
\centering
\includegraphics[width=.45\textwidth,height=\figH]{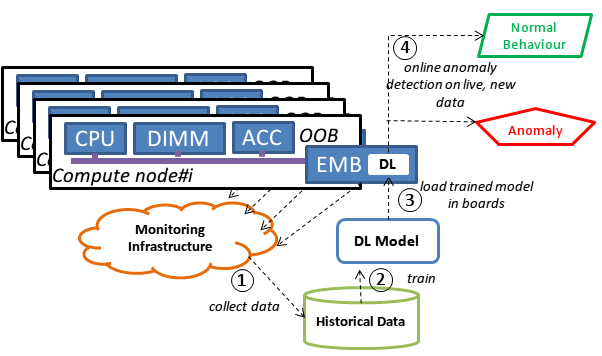}
\caption{Methodology Architecture}
\label{fig:methodology_scheme}
\end{figure}

In this paper we propose a system-oriented methodology to automatically detect
anomaly based on a ML model and relying on the data collected by a monitoring
infrastructure. The proposed scheme is depicted in
Figure~\ref{fig:methodology_scheme}. The supercomputer/data center nodes are
endowed with embedded boards that measure a set of fine-grained metrics
describing the system behaviour. Thanks to these measurements it is possible to
distinguish between anomalous and normal conditions. 

Data collected with the monitoring framework is fed to a ML model in order to
train it to detect anomalies. During the training phase the model is going to
encounter only examples describing a system under normal conditions. After the
training phase, the ML models are loaded on the embedded monitoring boards
(\emph{EMB} in the figure); when new measurements arrive, the trained ML model
takes them as input and can identify anomalies, triggering an alarm for system
admins. Since the embedded boards do not possess great computing capability, the
ML model must be lightweight and generate low overhead.

The benefits of edge computing (placing the anomaly detection module in the
embedded boards) is twofold: 1) the board can directly read the out-of-band
sensors measurements, reducing the access time (furthermore, in-band monitoring
is not allowed on many systems since it could affect stability); 2) online
inference could not be easily performed on the computing nodes of a HPC system
since it would subtract resources from users and complicate the scheduling
process.

\subsection{Target Supercomputer \& Monitoring Framework}
\label{sec:davide_examon}

As a test bed for our approach we used a supercomputer \cite{DAVIDE}.  which was
ranked in the late hundreds of the TOP500\cite{Top500} and among the twenty
first positions of the GREEN500\cite{Green500} in the November 2017 list.  It is
a supercomputer based on OpenPOWER platform and this allowed us the out-of-band
monitoring employed by our approach.  The system comprises 45 nodes connected
with Infiniband EDR \SI{100}{\giga \byte /\s} network, with a total peak
performance of \SI{990}{\tera Flops} and an estimated power consumption of less
than \SI{2}{\kilo \watt} per node. Each node is a 2 Open Unit (OU) Open Compute
Project (OCP) form factor and hosts two IBM POWER8 Processors with NVIDIA NVLink
and four Tesla P100 data center GPUs, with the intra-node communication layout
optimized for best performance.

On the target supercomputer we developed and deployed a lightweight and scalable
monitoring infrastructure, namely DiG (Dwarf in a
Giant)~\cite{DiG,DBLP:conf/cf/BartoliniBLBGTG18}.  Data coming from
heterogeneous sources is gathered with a high sampling rate and placed in common
storage area; this allow to easily retrieve historical data to be used for
training purposes. The collected data is stored in a distributed time series
database (KairosDB~\cite{kairosDB}), built on top of a NoSQL database Apache
Cassandra~\cite{cassandra2015apache} The framework comprises a set of agents
running on the edge of computing nodes, using embedded measuring boards. These
agents monitor the power consumption of each node at the plug as well as
performance and utilization metrics, using both software commands and hardware
sensors. The measured values are sent to a data management backbone (namely
ExaMon~\cite{beneventi2017continuous}), through a communication layer based on
the open-source MQTT (MQ Telemetry Transport) protocol~\cite{standard2014mqtt},
which is designed for low bandwidth, high latency networks and minimal resource
demands.

To measure the power consumption with high resolution, the monitoring agents
exploit a power sensing module and an embedded monitoring board (Beaglebone
Black - BBB~\cite{coley2013beaglebone}), one for each node. The BBBs are based
on an Arm Cortex-A8, and include a 12-bit ADC which allows sampling rates up to
50k samples per second and synchronization of the measurements within one
microsecond, thanks to the hardware support of the Precision Time Protocol
(PTP)~\cite{synchHPCS2016,synchANDARE2018}.  For an out-of-band monitoring of
the nodes performance we use the IBM Amester commands, which exploit the IPMI
interface\cite{slaight2003using} to the OpenPOWER POWER8 on-chip
controller\cite{rosedahl2015chip} (OCC), to get OCC sensor readings. The IPMI
Amester commands are sent to the OCC, through the Board Management Controller
(BMC), using a python script. The python script executes on the embedded
monitoring board. The granularity of the data is 5s and 10s, respectively for
IPMI metrics and OCC metrics. 

\section{Experimental Evaluation}
\label{sec:exp_evaluation}

As ML model to detect anomalies, we propose an approach based on a particular
type of neural network called \emph{autoencoder}\cite{goodfellow2016deep}. We
exploit the series of measurements (features) describing the state of the HPC
system or data center and collected with the ad-hoc monitoring infrastructure.
Under ``healthy'' operating conditions these features are connected by specific
relationships (i.e. the power consumption directly depends on the workload,
temperature is related to the clock frequency, etc.). These correlations are no
longer valid when the system enters an anomalous state. The main idea of our
method relies on the autoencoder capability to learn the typical (normal)
correlation between the measures and then consequently identify changes in this
correlation that indicate an abnormal situation. This research avenue  has been
partially explored in recent years, although not in the HPC
field\cite{lv2016fault,lee2017convolutional,costa2015fully,ince2016real,gabel2012latent}.
We teach the autoencoder the normal behaviour of the computing nodes; after the
training phase the autoencoder will be able to detect anomalous situations. 

How can an autoencoder detect abnormal conditions? An autoencoder is a neural
network trained to copy its input $I$ to its output $O$.  It has one or more
internal layers $H$ that try to the represent the data taken as input. An
autoencoder is split in two subparts: an encoding function $H = enc(I)$ and a
decoding function that reconstructs the input $O = dec(H)$. Typically,
autoencoders do not just learn the identity function $dec(enc(I))= I$ but are
designed not to create perfect copies, e.g. the dimension of the hidden layer
can be smaller than the dimension of the input. Thus, the output of an
autoencoder is generally different from its input and the difference between
input and output is called \emph{reconstruction error}.  

We take advantage of the reconstruction error in order to detect anomalies. We
train the autoencoder with data corresponding to the normal state and minimize
the reconstruction error; this error is called \emph{training error}. In this
way it learns the normal correlations between the features from the monitoring
infrastructure. After this first training phase, we feed the autoencoder with
new data unseen before -- this is generally called \emph{inference} in ML
terminology -- and we then observe the reconstruction error. If the new data is
similar to the data used as input (if it respects the normal correlations) then
the error will be small and comparable to the training error. If the new data
correspond to an anomalous situation, the autoencoder will struggle to perform
the reconstruction, since the learned correlations do not hold. Hence we
identify anomalies by observing large reconstruction errors (w.r.t. to the
training error) during the inference phase.

\subsection{Autoencoders Training}
\label{sec:AE_training}

We create an autoencoder for each node in the supercomputer since, as we will
describe in Sec.~\ref{sec:result}, dedicated models for each node outperform a
single, generic model to be applied to all nodes.  Each network is trained using
node related data collected by the monitoring infrastructure. In the training
phase, we use only data corresponding to the normal state.  The data set is a
collection of a couple of hundreds of metrics, ranging from core load, frequency
and temperature, to node power consumption, room temperature, GPUs usage,
cooling fans speed, etc. The metrics (also referred to as \emph{features}) form
the input set for the neural networks. 

For each node we have a training set corresponding to 2 months of normal
behaviour (obtained in collaboration with system admins). Due to storage reason,
the fine-grained monitoring data is not preserved for more than a week, hence
for the training we use coarse-grained data, where the measurements are
aggregated in five minute intervals. After collecting the raw data, we
pre-process it (for example we removed data corresponding to periods where the
monitoring system was not working properly) and normalize it. This preparation
takes about 30s. The final number of features is 166.

We adopted the same network topology for each node.  After an empirical
evaluation we chose a sparse autoencoder model\cite{boureau2008sparse}, that
proved to be the better option in terms of accuracy and computational demands,
especially for training/inference times. The network is composed of three fully
connected layers, an input layer, an output layer and one hidden layer. As
activation function for the neurons we use Rectified Linear
Units\cite{nair2010rectified} (ReLU); as regularization term (needed in sparse
networks) we employ the L1-norm loss
function\cite{alain2014regularized,goodfellow2016deep}. The input and the output
layers have as many neurons as the number of the features (166) while the
hidden-layer has ten times the number of features (thus 1660 in total).

Each autoencoder is trained with data from its corresponding node; for the
training we use the Tesla P100 GPUs available in the supercomputer nodes (each
node trains its own autoencoder). For the training phase we use use Adam
algorithm\cite{kingma2014adam}, with mean absolute error as target loss. After
some preliminary experiments we opted for a batch size equal to 32 and 100
epochs. The training takes around 20 seconds for each autoencoder. The training
times and overhead are not a critical concern since the training phase takes
place only once or with a very low rate (once every few months), and can be
scheduled during maintenance periods.  The framework used for the design and the
training of the autoencoders is Keras \cite{chollet2015keras} with TensorFlow
for GPU\cite{tensorflow2015-whitepaper} as a back-end.

\subsection{Online Inference}
\label{sec:AE_inference}

After training the autoencoders, we leverage the out-of-band monitoring system
and the embedded monitoring boards (BBB) to execute the inference online and
detect anomalies thanks to edge computing. We installed TensorFlow on the BBB
and took advantage of the NEON accelerator (SIMD architecture). On each BBB we
load the trained autoencoder of the corresponding node, then we feed it with new
data coming from the monitoring framework. The results of the detection are
presented in the following section. Here we want to point out that we process a
batch of input data (the set of 166 features) in just 11ms, which is a
negligible overhead considering the sampling rate of several seconds. 

\subsection{Results}
\label{sec:result}
We injected two different anomalies in the supercomputer used as test case: we
changed the frequency governor configuration of the computing nodes
(see\cite{brodowski2013cpu} for details). Changing the frequency governor
disrupts the relationships between core load and core frequency (and other
related features, such as power consumption, etc.). This is a misconfiguration
anomaly and should be detectable with our approach. The normal configuration is
\emph{default}; we changed it to \emph{powersave} and \emph{performance}, in
different periods of time\footnote{\emph{default}: the frequency of a core
directly depends on its load; \emph{powersave}/\emph{performance}: frequencies
are forced to the lowest/highest possible value}.  To evaluate the trained
autoencoders we consider the reconstruction error. As described in
Sec.~\ref{sec:approach} we expect to observe greater reconstruction errors
during anomalous periods w.r.t. normal ones.
 
\begin{figure}
\centering
\includegraphics[width=0.45\textwidth,height=\figH]{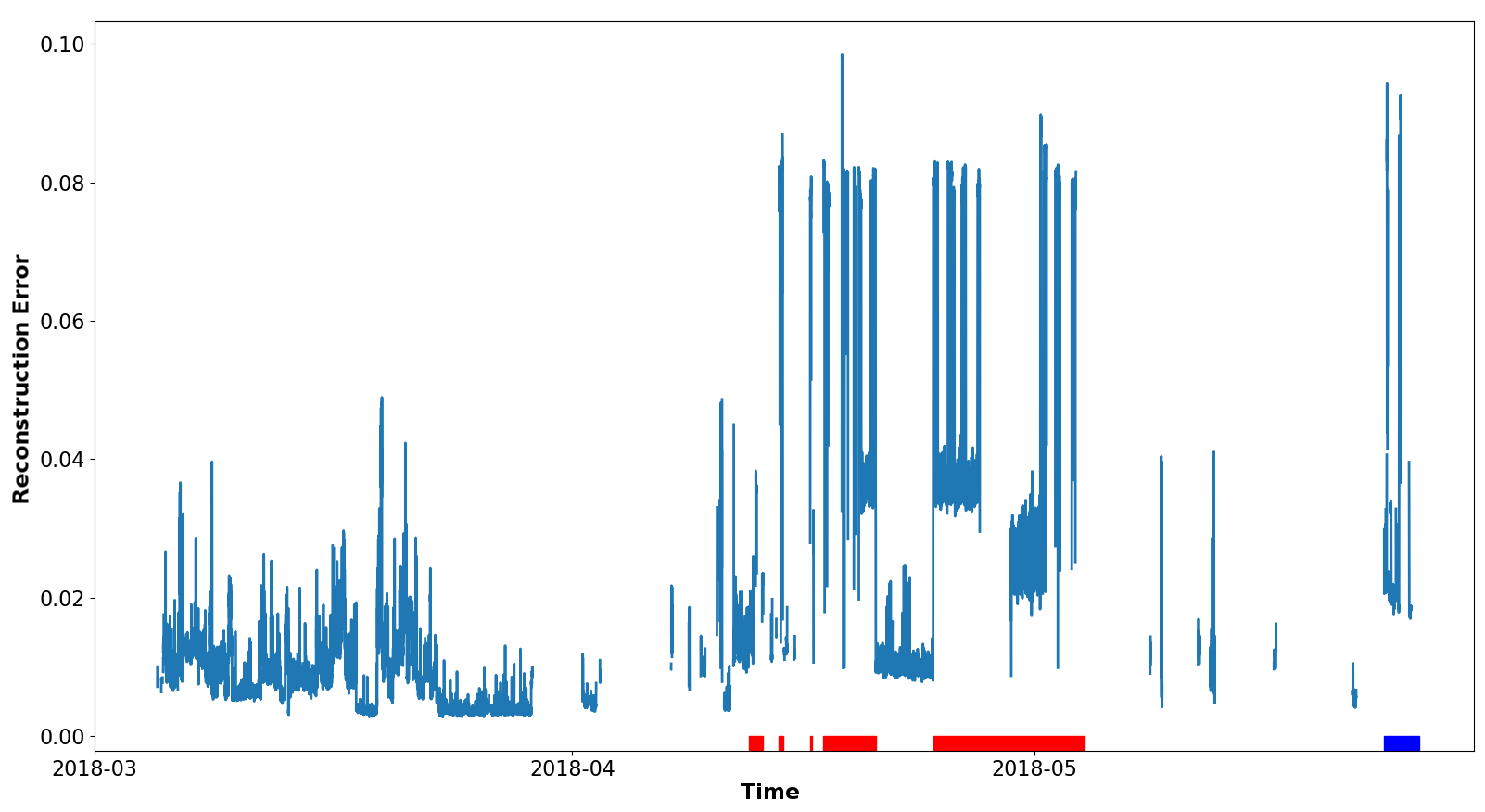}
\caption{Reconstruction error for node \emph{node45}}
\label{fig:recon_error_nodes_45}
\end{figure}

Our hypothesis is confirmed by quantitative and visual analysis of the
reconstruction error. Figure~\ref{fig:recon_error_nodes_45} plots the
reconstruction error ($y$-axis) obtained in a roughly two months period
($x$-axis), for a specific node. The reconstruction error trend is plotted with
a light blue line; the gaps in the line represent periods when the node was idle
and that have been removed from the data set (thus we exclude them from the
reconstruction error). We have 6 anomalous periods, identified by colored
highlights along the $x$-axis: during the first 5 (red lines) the frequency
governor was set to powersave, while during the last one (blue) the governor was
set to performance. It is possible to see that, on average, the reconstruction
error is definitely larger during anomalous periods compared to the normal ones.
This observation is confirmed by a quantitative analysis performed on all nodes:
the average normalized reconstruction error computed during normal periods
(excluding the training set) is equal to 1.08, while the average error obtained
during anomalies (grouping both types) is equal to 14.54. If we use a generic
autoencoder for all nodes (instead of a series of dedicated ones), the error for
the anomalous periods decreases to 6.29 while the error for normal periods
remains similar (1.01). To detect anomalies we prefer reconstruction errors as
large as possible with anomalous data, hence the set of dedicated models is
better than the generic one.

\textit{Accuracy Evaluation.} We use a threshold-based method to distinguish
anomalies from normal states. Suppose we have a data point $i$ that contains
features collected by the monitoring infrastructure. To classify it we feed it
to the trained autoencoder: if the reconstruction error $E_i$ is greater than a
threshold $\theta$, then the point is ``abnormal''; otherwise the data point is
considered normal. As threshold we choose the $n$-th percentile of the errors
distribution of the normal data set, where $n$ is a value that depends on the
specific autoencoder/node. To find the best $n$ for each autoencoder we employed
a simple generate-and-test search strategy, that is we performed experiments
with a finite number of values (after a preliminary empirical evaluation) and
then chose those guaranteeing the best results in term of classification
accuracy.  The fact that the best results are obtained by using a different $n$
for each node validates the choice of having multiple dedicated autoencoders
rather than a generic one.

The classification accuracy is measured by the
\emph{F-score}\cite{van1979information} for each class, \emph{normal} (N) and
\emph{anomaly} (A). F-score ranges between 0 and 1, with higher values
indicating higher accuracy. In Table~\ref{tab:classification_accuracy} we see
the results. In the first column there is the node name (we show the values for
a subgroup of nodes). The remaining columns report the F-score values for 3
different $n$-th percentiles; there are two F-score values for each $n$-th
percentile, one for the normal class (\emph{N}) and one for the anomaly class
(\emph{A}). We can see that the F-score values are very good, highlighting the
good accuracy of our approach, with an accuracy between 0.87 and 0.98. 

\begin{table}
\small\sf\centering
\begin{tabular}{ccccccc}
 \toprule
 \multirow{2}{*}{\emph{Node}} & \multicolumn{2}{c}{95-th perc.} & \multicolumn{2}{c}{97-th perc.} & \multicolumn{2}{c}{99-th perc.} \\
  & N & A & N & A & N & A \\ 
  \midrule
\emph{node17} & 0.97 & 0.89 & 0.98 & 0.93 & 0.99 & 0.97 \\  
\emph{node19} & 0.97 & 0.90 & 0.98 & 0.94 & 0.99 & 0.97 \\  
\emph{node45} & 0.97 & 0.92 & 0.98 & 0.95 & 0.99 & 0.98 \\ 
\emph{node29} & 0.97 & 0.75 & 0.98 & 0.82 & 0.99 & 0.85 \\ 
  \midrule
  \midrule
\emph{Average} & 0.96 & 0.87 & 0.98 & 0.91 & 0.97 & 0.89 \\
  \bottomrule
\end{tabular}
\caption{Classification Results}
\label{tab:classification_accuracy}	
\end{table}

\section{Conclusion}
\label{sec:conclusion}
In this manuscript we have presented an approach for automated anomaly detection
for large scale HPC environments and data centers. Our approach leverages
Machine Learning and edge computing for real-time anomaly detection. We use
autoencoders trained to learn the normal behavior of each computing node based
on its historical telemetry data of "good" behaviors. The autoencoders are
trained and tested on a in-production supercomputer and deployed as an extension
of the built-in embedded monitoring devices. 

\section*{Acknowledgment}
We want to thank CINECA and E4 for granting us the access to their systems. This
work has been partially supported by the EU H2020 FET project OPRECOMP (g.a.
732631)

\bibliographystyle{alpha}
\bibliography{bib}

\end{document}